\documentclass[sigconf]{acmart}

\AtBeginDocument{%
  }
\copyrightyear{2022}
\acmYear{2022}
\setcopyright{rightsretained}
\acmConference[MM '22]{Proceedings of the 30th ACM International Conference on Multimedia}{October 10--14, 2022}{Lisboa, Portugal}
\acmBooktitle{Proceedings of the 30th ACM International Conference on Multimedia (MM '22), October 10--14, 2022, Lisboa,Portugal}
\acmDOI{10.1145/3503161.3547743}
\acmISBN{978-1-4503-9203-7/22/10}

\begin{document}

\title{WOC: A Handy Webcam-based 3D Online Chatroom}



\iftrue
\author{Chuanhang Yan$^\dagger$$^\star$}
\affiliation{%
  \institution{Beijing Institute of Technology}
  \city{Beijing}
  \country{China}
}
\email{yanchxx@gmail.com}

\author{Yu Sun$^\dagger$$^\star$}
\affiliation{%
  \institution{Harbin Institute of Technology}
  \city{Harbin}
  \country{China}
}
\email{yusun@stu.hit.edu.cn}

\author{Qian Bao}
\affiliation{%
  \institution{JD Explore Academy}
  \city{Beijing}
  \country{China}
}
\email{baoqian@jd.com}

\author{Jinhui Pang$^\boxtimes$}
\affiliation{%
  \institution{Beijing Institute of Technology}
  \city{Beijing}
  \country{China}
}
\email{pangjinhui@bit.edu.cn}

\author{Wu Liu$^\boxtimes$}
\affiliation{%
  \institution{JD Explore Academy}
  \city{Beijing}
  \country{China}
}
\email{liuwu1@jd.com}

\author{Tao  Mei}
\affiliation{%
  \institution{JD Explore Academy}
  \city{Beijing}
  \country{China}
}
\email{tmei@live.com}
\fi


\thanks{$^\dagger$ Equal contribution. $^\boxtimes$ Corresponding author.} 
\thanks{$^\star$ This work was done when Chuanhang Yan and Yu Sun were interns at JD Explore Academy.} 
\renewcommand{\shortauthors}{Chuanhang Yan et al.}

\begin{abstract}
We develop WOC, a webcam-based 3D virtual online chatroom for multi-person interaction, which captures the 3D motion of users and drives their individual 3D virtual avatars in real-time. Compared to the existing wearable equipment-based solution, WOC offers convenient and low-cost 3D motion capture with a single camera. To promote the immersive chat experience, WOC provides high-fidelity virtual avatar manipulation, which also supports the user-defined characters. 
With the distributed data flow service, the system delivers highly synchronized motion and voice for all users.
Deployed on the website and no installation required, users can freely experience the virtual online chat at \textcolor{blue}{\textit{{https://yanch.cloud/}}}.
  
\end{abstract}



\maketitle

\vspace{-2mm}
\section{Introduction}

Metaverse-related techniques are drawing more and more attention in recent years. One of the most popular topics is to bring real-world interactions into the Metaverse, where remote users can interact face-to-face in the virtual world. To achieve this, we develop a webcam-based 3D virtual online chatroom, WOC. WOC integrates multi-source techniques such as monocular 3D motion capture, virtual avatar manipulation, auto rigging and animation for 3D characters, and distributed data flow service.

\begin{figure}
  \includegraphics[width=1\linewidth]{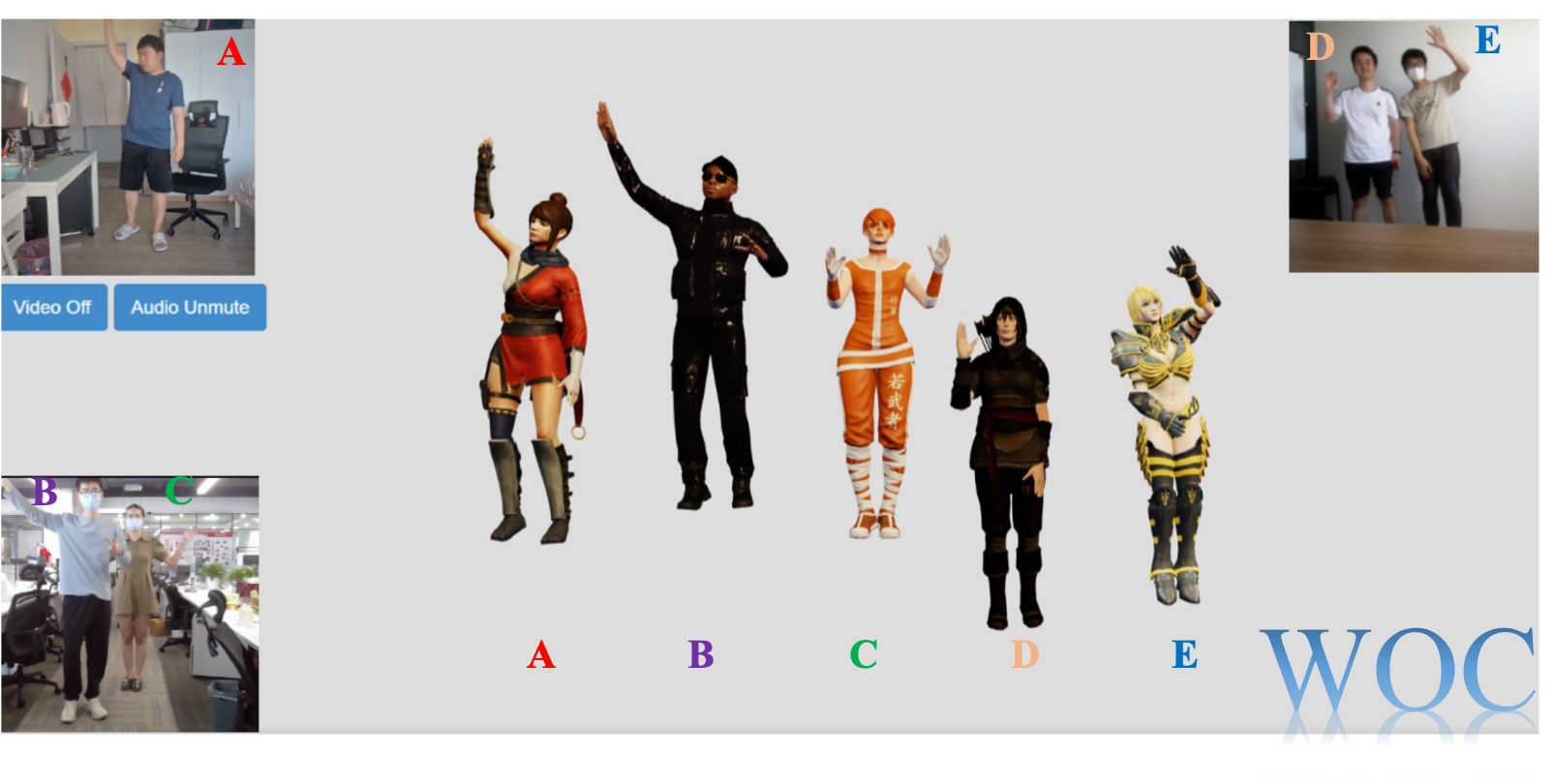}
  \vspace{-9mm}
  \caption{Illustration of our interactive chatroom WOC.}
  \vspace{-6mm}
  \label{fig:Demo}
\end{figure}

Different from the traditional 2D video chatroom, WOC captures the 3D body motion of users to manipulate their individual 3D virtual avatars for interaction in 3D space. Unlike wearable 3D motion capture devices (such as IMUs) that cost thousands of dollars and suffer from many limitations, WOC only requires a low-cost webcam, and no need to wear any equipment, allowing users to enjoy more flexibility and convenience, as shown in Fig.~\ref{fig:Demo}.

\begin{figure*}
  \includegraphics[width=0.9\textwidth]{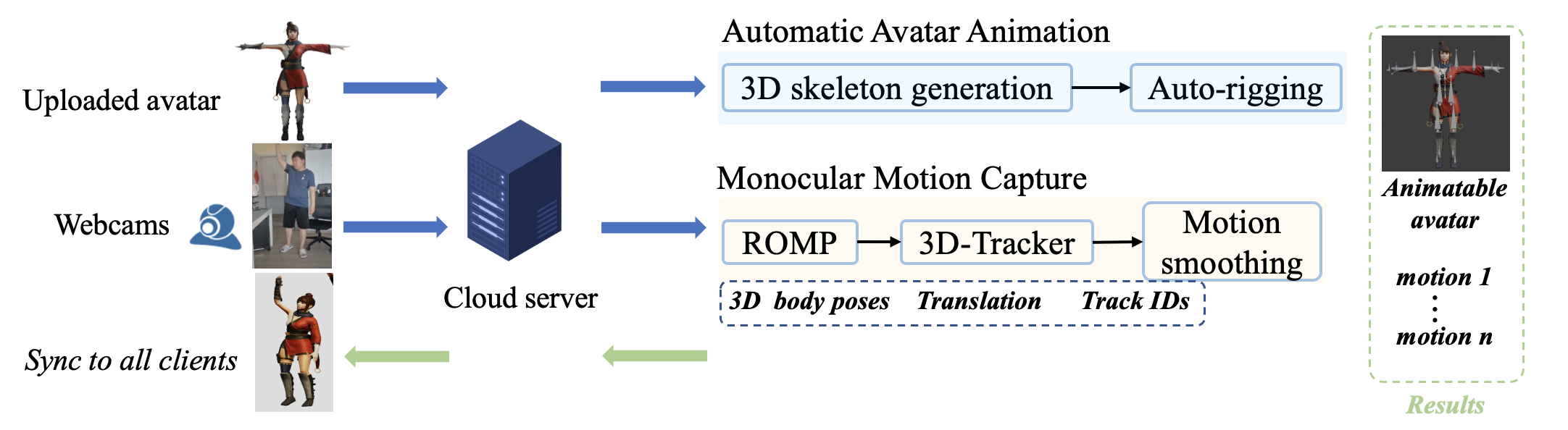}
  \vspace{-5mm}
  \caption{An overview of WOC. The main modules include monocular motion capture, automatic avatar animation, and distributed deployment and visualization. }
  \label{fig:framework}
  \vspace{-5mm}
\end{figure*}

\textbf{How to use WOC?}
WOC is deployed on web pages for easy access by users anytime and anywhere. 
Before joining the chatroom, users need to pick up their preferences from the default avatars or upload their own avatars.
Then they just need to click to open their webcams. 
WOC would put the (multiple) people presented in front of the webcams into a chatroom. 
One webcam can capture the motion of multiple people. 
Users would be able to drive their individual virtual avatars to interact with others in real-time. 

\vspace{-3mm}
\section{System Architecture}
Our goal is to provide a lightweight, flexible, and real-time multi-person virtual chatroom for users. We design a distributed architecture powered by the cloud service to handle the multi-source computations. As shown in Fig.~\ref{fig:framework}, the system mainly consists of three modules. The system simultaneously receives videos from each user's webcam, and feeds them to the monocular motion capture module. The output 3D motions are responded by the automatic avatar animation module that automatically transfers the motions to each corresponding avatar picked/uploaded by the users. 
A cloud-based service integrates all the modules, encodes the data flow, and sends the real-time interactive chat back to the users.

\vspace{-2mm}
\subsection{Monocular Motion Capture}
To meet the need for online real-time chat, we need to simultaneously estimate the 3D motion states~\cite{liu2021recent} of all people presented in the webcams. 
Built on the real-time multi-person motion capture method, ROMP~\cite{ROMP}, we estimate 3D motion in three steps.

\textbf{3D Motion Capture from a Single Image.} 
Given a single RGB image, ROMP estimates the 3D motion state of each person, represented as 3D SMPL~\cite{loper2015smpl} pose parameters and 3D translation $\boldsymbol{T} \in \mathbb{R}^{3}$ in camera space.
However, ROMP was trained to estimate from a single image. And its video predictions suffer from temporal jittering. 
To extract the smooth 3D body motion sequence of all the people presented in the webcam, we further need to
associate the predictions via tracking and motion smoothing.



\textbf{3D-detection-based Tracking.} 
Based on the popular 2D tracker ByteTrack~\cite{zhang2021bytetrack,liu2018t}, we develop a 3D tracking algorithm 3D-Tracker, which takes the distinguishable 3D translation of ROMP as input to replace the original 2D bounding box input. 
In this way, we do not need to involve an additional 2D detector to perform tracking, which greatly alleviates the computational burden on our real-time system. 
More specifically, we convert ROMP's predicted 3D translation $\boldsymbol{T}=(x,y,z)$ to $(u,v,z,s)$ where $(u,v)$ are the 2D image coordinates obtained via projecting 3D position to 2D image plane, $s$ is the inverse of $z$ to present the person scale in image. 
Therefore, the system can associate the single-frame predictions to form the 3D body motion sequence of each person. 

\textbf{Motion Smoothing.} 
To filter out the high-frequency jittery in the predicted motion sequence, we develop a motion smoothing sub-module. This module is built on the well-known OneEuro filter~\cite{casiez2012}.
For each 3D body motion sequence, we create three individual OneEuro filters to separately process the estimated body 3D orientations, 3D poses, and 3D translations, and then obtain the stable and accurate 3D body motion sequence of all people presented in the webcam frames for avatar animation.  
\vspace{-3mm}

\subsection{Automatic Avatar Animation}
To animate the 3D avatars uploaded by users with estimated SMPL pose parameters, we develop an auto-rigging blender addon\footnote{https://github.com/yanchxx/CDBA} to generate and bind compatible SMPL skeletons to 3D avatars.

\textbf{3D SMPL Skeleton Generation.} 
Given a new uploaded 3D avatar, we first normalize the scale of its mesh and align it to the origin.
Then we employ a MeshCNN from NBS~\cite{NBS} to extract the geometric and topological features from the normalized mesh and estimate the 3D position of 24 joints defined by SMPL.

\textbf{Automatic Rigging.} 
To bind the generated 3D skeleton to the avatar, we employ an auto-rigging algorithm embedded in Blender~\cite{Blender} to calculate the skinning weights of each joint. 
In this way, we can convert the joint 3D rotation, represented by estimated SMPL pose parameters, into the motion of the mesh vertices. 
Besides, for Maximo\footnote{https://www.mixamo.com/} users, we also develop a module to convert the avatar to our format for further animation. 
\vspace{-2mm}

\vspace{-1mm}
\subsection{Distributed Deployment}
To alleviate the computational burden on users' clients and enable flexible usage, WOC is deployed in a distributed architecture. On user client, with permission from users, video and audio are collected from their webcams and microphones.
To send the data stream to the cloud server, the system establishes peer-to-peer communication via WebRTC.  
From the cloud server, the system receives the motion sequences of all people for animating  avatars.
We employ the three.js to show the generated 3D avatar animations on the webpage. 

The cloud server collects the media streaming and the uploaded new avatars from all user clients. 
For each user client, we launch an individual process of motion capture module to estimate the 3D body motion from collected video frames. 
Then the synchronized results are sent back to all users within the same chatroom.
In this way, every user would be able to chat with each other in the chatroom, as shown in Fig.~\ref{fig:Demo}.
\vspace{-3mm}


\section{System Performance}
WOC collects webcam frames of size 512$\times$512 from each user client, which are not saved anywhere to avoid the invasion of user privacy.
We test the deployment of the back-end of WOC on different servers. 
A server with a single 1070Ti/3090Ti GPU can achieve over 30/60 FPS, which means it can support the real-time online 3D chat from two/four clients. 
A server with eight 3090 GPUs, we launch individual processes on different GPUs for each user and ensure running at 20 FPS for all users. 
Without considering the network delay, on the cloud server, the latency is about 50ms from receiving the image to sending the result.

\begin{acks}
This work was supported by the National Key R\&D Program of China under Grant No.2020AAA0108600.
\end{acks}

\bibliographystyle{ACM-Reference-Format}
\bibliography{references}

\end{document}